\documentclass[twocolumn,showpacs,preprintnumbers,nofootinbib]{revtex4}
\usepackage{graphicx}
\usepackage{latexsym}
\usepackage{amssymb}

\def\lsim{\mathrel{\rlap{\lower4pt\hbox{\hskip1pt$\sim$}}
    \raise1pt\hbox{$<$}}}
\def\gsim{\mathrel{\rlap{\lower4pt\hbox{\hskip1pt$\sim$}}
    \raise1pt\hbox{$>$}}}
\def\sqr#1#2{{\vcenter{\vbox{\hrule height.#2pt
         \hbox{\vrule width.#2pt height#1pt \kern#1pt
         \vrule width.#2pt}
         \hrule height.#2pt}}}}

\def\beq{\begin{equation}}
\def\eeq{\end{equation}}
\def\beqa{\begin{eqnarray}}
\def\eeqa{\end{eqnarray}}
\begin{document}

\title{Dark energy and the Rutherford-Soddy radiative decay law}

\author{M. C. Bento}
\altaffiliation[ ] {Also at Centro de F\'isica Te\'orica e de Part\'iculas, IST. Email
address: bento@sirius.ist.utl.pt}

\author{O. Bertolami}
\altaffiliation[ ] {Also at Instituto de F\'isica dos Plasmas, IST.
Email address: orfeu@cosmos.ist.utl.pt}

\affiliation{Instituto Superior T\'ecnico, Departamento de F\'isica \\
Av. Rovisco Pais 1, 1049-001 Lisboa, Portugal}

\vskip 0.5cm

\date{\today}

\begin{abstract}

It is shown that a putative evolution of the fundamental
couplings of strong and weak interactions via coupling to dark energy
through a generalized Bekenstein-type model may, for a linear model of
variation,
cause deviations on the statistical nuclear decay Rutherford-Soddy law
unless bounds are imposed on the parameters of this variation.
Existing bounds for the weak interaction exclude any significant deviation. Bounds
on the strong interaction are much less stringent.

\vskip 0.5cm

\end{abstract}

\pacs{98.80.-k,98.80.Cq,12.60.-i\hspace{4cm} Preprint DF/IST-9.2008}

\maketitle
\vskip 2pc
\section{Introduction}

Recently reported evidence of a putative time variation of the electromagnetic
coupling on cosmological time scales from the observation of absorption spectra
of quasars \cite{Webb} has led to a revival of interest on ideas that actually go back to Dirac
\cite{Dirac} about the variation of fundamental couplings (see Ref. \cite{Uzan}
for a review). Such variations are fairly natural in unification models with extra dimensions
or expectation values of scalar fields. It is likely that if a variation of the electromagnetic coupling
occurs one should also expect variations of other gauge or Yukawa couplings.
Confirming these observations and proposing
realistic particle physics models consistent with a cosmological variation of the fundamental coupling
constants is therefore of great interest. In this work we shall consider the effect that a variation
of the strong and weak couplings may have on the nuclear decay Rutherford-Soddy law which
is quite compatible with the decay of short lived nuclei as $^{238}Pu$ \cite{Cooper} . For that we
shall consider a generalized version of the Bekenstein model \cite{Bekenstein} and the available
bounds on the variation of those couplings.

A relevant feature of the observed variation of the electromagnetic coupling  is that is is presumably a late
event in the history of the universe and it is thus natural to associate it to the observed late
accelerated expansion of the universe \cite{Perlmutter}. This late time acceleration of the
expansion can be driven by a cosmological constant (see \cite{Bento1} and
references therein), by a scalar field with a suitable potential, quintessence \cite{Ratra},
or by a fluid with an exotic equation
of state like the generalized Chaplygin gas (GCG) \cite{Kamenshchik}. We mention that for
the evolution of the fine structure constant along the lines of the Bekenstein
model \cite{Bekenstein} and generalizations, a cosmological constant has been
suggested as driving entity \cite{Equivalence} as well as a ultra-light scalar field \cite{Gardner}
and a dynamical dark energy field  \cite{Anchordoqui,Bento04a,Bertolami04,dent}.
In this work we shall consider the variation of the strong
and weal couplings in the context of a exponential quintessence model and of the GCG model.

\section{Nucleon decay and the evolution of the strong and weak couplings}

It is a quite well established experimental fact that the nuclear decay rate
of any nuclei is described by the statistical Rutherford-Soddy law,
first stated back in 1902:
\beq
{dN\over dt} =-\lambda N~~,
\eeq
where $\lambda={1\over \tau_0}$, $\tau_0$ being the nuclei's lifetime.
For a constant $\lambda$, which reflects, as later stated by Rutherford,
Chadwick and Ellis, the fact that {\it the rate of transformation
of an element has been found to be constant under all conditions}
\cite{RCE30}, one obtains the
well known exponential decay law:
\beq
N(t)=N_i e^{-\lambda (t-t_i)}~~,
\eeq
where $N_i$ is the nuclei's number counting at $t_i$.
This statistical law reflects radiative processes of electromagnetic,
strong and weak nature that take place within the nuclei. Thus,
$\tau_0$ is related to the amplitude of the relevant decay process being
therefore a function of order $\gamma$ of the coupling constant $\alpha_g$:
\beq
\tau_0=A \alpha_g^\gamma~~,
\eeq
with $\alpha_S(E) = g_s^2(E)/\hbar c$ for the strong interactions, $g_s$ being the strong coupling, and
$\alpha_W=G_F  m_p^2 c/\hbar^3$ for the weak interactions, $G_F$
being Fermi's constant and $m_p$ the proton mass. Parameter A is a constant related to a specific
process, suitable integration over the phase space, binding energy, quark masses, etc.

On the other hand, one expects that if dark energy couples with the whole
gauge sector of the Standard Model, this coupling can be modelled by
a generalization of the
so-called Bekenstein model \cite{Bekenstein}
\beq
{\cal L}_{gauge} =-\frac{1}{16\pi} f\left(\phi\right)F_{\mu\nu}^a F^{\mu\nu a}~,
\eeq
where $f$ is an arbitrary function of the dark energy field, $\phi$, and $F_{\mu\nu}^a$ the gauge field strength.
Given that the variation of the
gauge couplings is presumably small (c.f. below), we expand this function to first order

\beq
f\left(\phi\right)=\frac{1}{\alpha_{i}}\left[1+\zeta \left({\phi-\phi_i \over M}\right)\right]~,
\eeq
where $\alpha_{i}$ is the initial value of the gauge structure constant, $\xi$ is a constant,
$\phi_i$ is the initial value of the quintessence field
and $M$ a characteristic mass scale of the dark energy model.
A model with a quadratic variation is discussed in Ref.
\cite{OP08}. It follows that
\beq
\left[f\left(\phi\right)\right]^{-1} \simeq\alpha_{i}\left[1-\zeta
\left({\phi-\phi_i \over M}\right)\right]~.
\label{eq: functionf}
\eeq

In this case, the gauge coupling, $\alpha_g$, evolution is given by
\beq \
\alpha_g =\left[f\left(\phi\right)\right]^{-1} \simeq \alpha_{i}\left[1-\zeta_g
\left({\phi-\phi_i \over M}\right)\right]~,
\label{alpha2}
\eeq
\noindent
hence, for its variation, we obtain
\beq {\Delta\alpha_g \over \alpha_g}=\zeta_{g}\left({\phi-\phi_i \over M}\right)~.
\label{delta2}
\eeq

For the electromagnetic interaction one should take into account the
Equivalence Principle limits, which implies \cite{Equivalence}
\beq
\zeta_{EM} \leq 7 \times 10^{-4}~.
\label{zetaF}
\eeq
\noindent
Of course, this bound does not apply to
short range interactions. However, as we shall see, this parameter must be
constrained in order to ensure that strong and weak gauge coupling do not change significantly
so to ensure the experimental standing of the Rutherford-Soddy law.

In what follows we consider only strong and weak decays given that the
effects of a putative variation of the electromagnetic coupling have been the subject
of various studies (see e.g. Ref. \cite{Bento04a} and references therein).
Thus, for strong and weak decays, keeping quark masses unchanged and
disregarding the running of the strong interactions, one can write for $\tau$:
\beq
\tau_{DE}=A f(\phi)^{-\gamma}~~,
\eeq
and since $\zeta < 1$
\beq
\tau_{DE}\simeq \tau_0\left[1-\gamma \zeta_{S,W} \left({\phi_0-\phi_i \over M}\right)\right]~~,
\eeq
depending on whether the process is ruled by strong or weak interactions,
where $\phi_0$ refers to the value of the dark energy scalar field at present.
Introducing this expression into the rate of change of atoms
\beq
{d N \over dt}\simeq -{N\over \tau_0}\left[1+\gamma \zeta_{S,W} \left({\phi_0-\phi_i \over M}\right)\right]~~,
\eeq
from which follows that
\beq
N(t)\simeq N_i \exp\left[-\lambda(\Delta t +\gamma \zeta_{S,W} I)\right]~~,
\eeq
where
\beq
I=\int_{t_i}^{t_0} \left({\phi_0-\phi_i \over M}\right) dt~~,
\label{integral}
\eeq
and $\Delta t=t_0-t_i$.

We now consider the bounds on the variation of the gauge couplings
(latest bounds on the variation of the electromagnetic coupling
can be found for instance in \cite{Bento04a} and references therein).
Regarding the strong interaction, considerations on the
stability of two-nucleon systems yield the bound \cite{Uzan}
\beq
\left|{\Delta \alpha_S \over \alpha_S}\right|< 4\times 10^{-2}~~,
\label{boundstrong}
\eeq
for the deuteron, and
\beq
\left|{\Delta \alpha_S \over \alpha_S}\right|< 6\times 10^{-1}~~,
\eeq
for the diproton.

For the weak interaction, one finds, from Oklo data \cite{Uzan}
\beq
\left|{\Delta \alpha_W \over \alpha_W}\right|< 4\times 10^{-3}~~.
\label{boundweak}
\eeq
On the other hand, from the weak interaction contribution to the nuclear ground
state energy of $^{147}Sm$, one gets \cite{Uzan}
\beq
\left|{\Delta \alpha_W \over \alpha_W}\right|< 0.02~~.
\eeq

From considerations on the $\beta$-decay rates of $^{187}Re$ lead to
\cite{re,dent}
\beq
{\Delta \alpha_W \over \alpha_W} < 3 \times 10^{-7}~~.
\eeq

\section{Dark Energy Models}

In order to quantify our predictions concerning the level of validity of the
statistical Rutherford-Soddy decay law, we consider two observationally
viable dark energy models.

\subsection{A quintessence model with an exponential potential}

For simplicity let us start considering the exponential potential
$V(\phi)=V_o e^{-\beta\phi /M_P }$, in the case of scalar field
domination, there is  a family of solutions \cite{Ferreira}
\beqa
\phi(t) &= &\phi_0+ \frac{2}{\beta} \ln (t M_P)\\
\qquad \phi_0 &=& \frac{2}{\beta} \ln\left(\frac {V_o\beta^2 }{2M_P^4(6-\beta^2)} \right)\\
\qquad \rho_\phi &\propto & \frac{1}{a^{\beta^2}}
\qquad a \propto t^{2/\beta^2}
\label{eq: jhsoln}
\eeqa
where $a$ is the scale factor of the universe. For $\beta < \sqrt{6}$, and $\phi_0$
can always be chosen
to be zero by redefining the origin of $\phi$, in which case
$V_0=\frac{2}{\beta^2}(\frac{6}{\beta^2}-1)M_P^4$. The parameters of
this model can be set to satisfy all
phenomenological constraints (see e.g. last reference in \cite{Ratra}).

\subsection{The Chaplygin gas model}

The GCG model considers an exotic perfect fluid described by the equation
of state \cite{Kamenshchik}
\beq p = -\frac{A}{\rho ^{\alpha}_{ch}}~, \label{GCGes}
\eeq
\noindent where
$A$ is a positive constant and $\alpha$ is a constant in the range
$0\leq \alpha \leq 1$. For $\alpha=1$, the
equation of state is reduced to the Chaplygin gas scenario.
The covariant conservation of the energy-momentum tensor for an homogeneous
and isotropic spacetime implies that
\beq
\rho_{ch} = \rho_{ch0}\Bigl[ A_s +
\frac{(1-A_s)}{a^{3(1+\alpha)}}\Bigr]^{\frac{1}{1+\alpha}}~,
\label{GCGrho}
\eeq
where $A_s=A/\rho_{ch0}^{(1+\alpha)}$ and
$\rho_{ch0}$ is the present energy density of the GCG.
Hence, we see that at early times
the energy density behaves as matter while at late times
it behaves like a cosmological constant. This dual role
is at the heart of the surprising properties of the GCG
model. Moreover, this dependence with the scale factor
indicates that the GCG model can be interpreted as an
entangled admixture of dark matter and dark energy.

This model has been thoroughly scrutinized from
the observational point of view and is shown to be compatible
with the Cosmic Microwave Background Radiation
(CMBR) peak location and amplitudes \cite{Bento03,Barreiro08}, with SNe
Ia data \cite{Alcaniz,realGCG,Bento05}, gravitational lensing
statistics \cite{Silva},
cosmic topology \cite{Bento06}, gamma-ray bursts \cite{BS06} and variation
of the electromagnetic coupling \cite{Bento07}.
The issue of structure
formation and its difficulties \cite{Sandvik} have been
recently addressed \cite{Bento04b}. Most recent analysis based on CMBR data
indicates that $\alpha < 0.25$ and
$A_s > 0.93 $ \cite{Barreiro08}.

Following Ref. \cite{realGCG,Bernardini08}, we describe the GCG through a real scalar
field. We start with the Lagrangian density
\beq
{\cal L}=\frac{1}{2}\dot\phi^2-V(\phi)~,
\label{realL}
\eeq
\noindent
where the potential for a flat, homogeneous and isotropic universe has the following form \cite{Bernardini08}
\beq
V={1\over 2} A_s^{1\over 1+\alpha}\rho_0\Bigl[\cosh(\beta \phi)^{\frac{2}{\alpha+1}}
+\cosh({\beta\phi})^{-\frac{2\alpha}{\alpha+1}}\Bigr]~,
\label{realV}
\eeq
\noindent
where $\beta=3(\alpha+1)/2$.

For the energy density of the field we have
\beq
\rho_\phi=\frac{1}{2}\dot\phi^2+V(\phi)=\Bigl[ A +
\frac{B}{a^{3(1+\alpha)}}\Bigr]^{\frac{1}{1+\alpha}}=\rho_{ch}~.
\label{realRHO}
\eeq

The Friedmann equation can be written as
\beq
H^2={ 8 \pi G \over 3} \rho_{\phi}
\eeq
and thus
\beq
H(t)=H_0 \Omega_\phi^{1/2}
\label{H}
\eeq
where $\Omega_\phi$ can be written as
\beq
\Omega_\phi=\left[A_s+{(1-A_s)\over a^{3(1+\alpha)}}\right]^{1\over 3(1+\alpha)}~.
\eeq

Using Eq.~(\ref{GCGes}) to obtain the pressure and integrating we find \cite{Bernardini08}
\beq
\phi (a)=-\frac{1}{2\beta}\ln\left[{\sqrt{1-A_s(1-a^{2\beta})}-\sqrt{1-A_s}\over
\sqrt{1-A_s(1-a^{2\beta})} + \sqrt{1-A_s}}\right]~.
\label{integratedPHI}
\eeq

\section{Results}

We have computed the integral $I$, Eq. (\ref{integral}), for the quintessence
model with an exponential potential (Model I) and
for the GCG model (Model II) and considered the implications
for the variation of the Rutherford-Soddy decay law for
strong and weak interactions. For Model I we have set $\beta^2 = 2$ and that
at present the scalar field contribution to the energy density
is $70\%$ of the critical density. For
Model II, we have considered the values $\alpha =0.2$, $A_s= 0.95$, and used
the relationship $1+z=1/a$ which
allows one to express the integral $I$ in terms of z.

For the strong interaction decay we have considered the $^{238}_{92}U$
nucleus for which $\tau_0= 4.468 \times 10^9$ years. For the
weak interaction decay we have considered the  $\beta^-$ decay of $^{187}Re$ for
which $\tau_0= 4.558 \times 10^9$ years. Using bounds (\ref{boundstrong}), (\ref{boundweak}),
and the integration interval equal to $\tau_0$
we compute $\zeta_{S,W}$ and the respective deviation from the Rutherford-Soddy law  $N/N_{RS}$
for the case of weak and strong
interactions with no coupling to dark energy. Our results are depicted in Table 1 for $\gamma=2$.
\begin{table*}[ht]
\centering
\begin{tabular}{c c c c c c c c}
\hline
\hline
& I (yrs) & $\mid \zeta_S\mid$ &$\mid\zeta_W\mid$ &
 $\mid{N\over N_{RS}}\mid_S -1\mid$ & $\mid{N\over N_{RS}}\mid_W -1\mid$\\
\hline
\\
Model I  & $1.22\times 10^{9}$ & $  7.1\times 10^{-2}$& $5.24 \times 10^{-7}$&
$3.8\times 10^{-2}$ & $2.79 \times 10^{-7}$  \\
\\
Model II & $1.08\times 10^{9}$ & $2.27\times 10^{-1} $& $1.7 \times 10^{-6} $&
$1.04 \times 10^{-1}$&  $8.1 \times 10^{-7}$
   \\
\\
\hline
\end{tabular}
\end{table*}

\section{Discussion and Conclusions}

In this work we have considered the impact that the coupling of dark energy
to gauge fields may have on the nuclear decay Rutherford-Soddy law. From the
available bounds on the variation of the strong and weak gauge couplings we
have estimated, in a linear model for the couplings, the
deviations from the Rutherford-Soddy law. We find that thanks to the stringent bound on
the variation of the weak coupling arising from $^{187}Re$ imply in significant constraints on any
variation of the Rutherford-Soddy law for weak interactions, actually at the level $3 \times 10^{-7}$ for
the quintessence field with an exponential potential and at $8 \times 10^{-7}$ level for the GCG model.
For the case of strong interactions the bounds are much more modest and are
about $4 \times 10^{-2}$ for the quintessence
model with exponential potential and $10^{-1}$ for the GCG model.
More stringent
bounds on the deviation of the nuclear decay law for strong interactions could be obtained if tighter
bounds on the variation of couplings were known. Furthermore, it is
interesting that rather different deviations are found for distinct
dark energy models. This may be relevant to distinguish them given that most
often they yield (or can be fiddled) to give rise to rather similar behaviour
in what concerns the cosmological observable parameters.

Our results indicate that, at least for strong interactions, bounds on the coupling constant
are not constrained enough or that dark energy is not
coupled to gauge fields, at least not through a linear variation model.

\begin{acknowledgments}

This work  is partially supported by the Funda\c c\~ao para a Ci\^encia e a
Tecnologia (FCT) project POCI/FIS/56093/2004.

\end{acknowledgments}



\begin{thebibliography}{99}

\bibitem{Webb} J.K. Webb et al., Phys. Rev. Lett. 82, 884 (1999); 87, 091301 (2001);
M.T. Murphy, J.K. Webb, V.V. Flambaum and S.J. Curran, Astrophys. Space Sci, 283, 577 (2003).

\bibitem{Dirac} P.M.A. Dirac, Nature, 139, 323 (1937).

\bibitem{Uzan} J.P. Uzan, Rev. Mod. Phys. 75, 403 (2003).

\bibitem{Cooper} P.S. Cooper, arXiv: 0809.4248.

\bibitem{Bekenstein} J.D. Bekenstein, Phys. Rev. D 25, 1527 (1982).

\bibitem{Perlmutter} S.J. Perlmutter  et al.  (Supernova Cosmology
Project), Ap. J. 483, 565 (1997); Nature, 391, 51 (1998);
A.G.  Riess  et al., (Supernova Search Team), Astron. J. 116, 1009 (1998);
P.M. Garnavich  et al., Ap. J. 509, 74 (1998); J.L. Tonry et al. [Supernova Search Team Collaboration], Ap. J. 594, 1 (2003).

\bibitem{Bento1}  M.C. Bento, O. Bertolami,  Gen. Rel. Gravitation 31, 1461 (1999);
M.C. Bento, O. Bertolami, P.T. Silva, Phys. Lett. B 498, 62 (2001).

\bibitem{Ratra} M. Bronstein, Phys. Zeit. Sowejt Union 3, 73
(1993);
O.  Bertolami, Nuovo Cimento 93B, 36 (1986); Forch. Phys.
34, 829 (1986);
M. Ozer, M.O. Taha, Nucl. Phys. B 287, 776 (1987);
B. Ratra, P.J.E.  Peebles, Phys. Rev. D 37, 3406 (1988); Ap. J. Lett. 325, 117 (1988);
C. Wetterich, Nucl. Phys. B 302, 668 (1988);
 R.R.  Caldwell, R.  Dave, P.J.  Steinhardt, Phys. Rev. Lett. 80, 1582 (1998);
I. Zlatev, L. Wang, P.J. Steinhardt, Phys. Rev. Lett. 82, 986 (1999);
J.P. Uzan, Phys. Rev. D 59, 123510 (1999);
O. Bertolami, P.J. Martins, Phys. Rev. D 61, 064007 (2000); Y. Fujii,  Phys. Rev. D 61, 023504 (2000);
A.A. Sen, S. Sen, S. Sethi, Phys. Rev. D 63, 107501 (2001);
M.C. Bento, O. Bertolami, N.C. Santos, Phys. Rev. D 65, 067301 (2002).

\bibitem{Kamenshchik} A. Kamenshchik, U. Moshella and V. Pasquier, Phys.
Lett. B 511, 265 (2001); N. Bili\'c, G.B. Tupper and R.D. Viollier, Phys. Lett. B
535, 17 (2002); M.C. Bento, O. Bertolami and
A.A. Sen, Phys. Rev. D 235, 043507 (2002).

\bibitem{Equivalence} K.A. Olive and M. Pospelov, Phys. Rev. D 65, 085044
(2002).

\bibitem{Gardner} C.L. Gardner, Phys.Rev. D 68, 043513 (2003).

\bibitem{Anchordoqui} L. Anchordoqui and H. Goldberg, Phys. Rev. D 68, 083513 (2003).

\bibitem{Bento04a} M.C. Bento, O. Bertolami, N.M.C. Santos, Phys. Rev. D 70, 107304 (2004).

\bibitem{re} K.A. Olive, M. Pospelov, Y.-Z. Qian, G. Manhes,
E. Vangioni-Flam, A. Coc, M. Casse, Phys. Rev. D 69, 027701 (2004);
Y. Fujii, A. Iwamoto, Phys. Rev. Lett. 91, 261101 (2003).

\bibitem{dent} T. Dent, S. Stern, C. Wetterich, arXiv:0809.4628 [hep-ph].

\bibitem{Bertolami04} O. Bertolami, R. Lehnert, R. Potting, A. Ribeiro, Phys. Rev. D 69,083513 (2004).

\bibitem{RCE30} S.E. Rutherford, J. Chadwick and C. Ellis, {\it Radiations
from Radioactive Substances} (Cambridge University Press 1930).

\bibitem{OP08} K.A. Olive and M. Pospelov, Phys. Rev. D 77, 043524
(2008).

\bibitem{Ferreira} P.G. Ferreira and M. Joyce, Phys. Rev. D58, 023503 (1998).

\bibitem{Bento03} M.C. Bento, O. Bertolami and A.A. Sen, Phys. Rev. D
67, 063003 (2003); Phys. Lett. B 575, 172 (2003); Gen.
Relativity and Gravitation 35, 2063 (2003).

\bibitem{Barreiro08}T. Barreiro, O. Bertolami and
P. Torres, Phys. Rev. D 78, 043530 (2008).

\bibitem{Alcaniz} J.S. Alcaniz, D. Jain and A. Dev,
Phys. Rev. D 67, 043514 (2003).

\bibitem{realGCG} O. Bertolami, A.A. Sen, S. Sen and P.T. Silva, Mon. Not.
Roy. Astron. Soc. 353, 329 (2004).

\bibitem{Bento05} M.C. Bento, O. Bertolami, N.M.C. Santos and A.A. Sen, Phys. Rev.
D71, 063501 (2005).

\bibitem{Silva} P.T. Silva and O. Bertolami, Ap. J. 599, 829 (2003);
A. Dev, D. Jain, D.D. Upadhyaya and J.S. Alcaniz, Astron. Astrophys. 417,
847 (2004).

\bibitem{Bento06} M.C. Bento, O. Bertolami, M.J. Rebou\c cas and P.T. Silva, Phys. Rev. D 73, 043504(2006).

\bibitem{BS06} O. Bertolami and P.T. Silva, Mon. Not. Roy. Astron. Soc. 365, 1149 (2006).

\bibitem{Bento07} M.C. Bento, O. Bertolami and P. Torres, Phys. Lett. B648, 14 (2007).

\bibitem{Sandvik} H. Sandvik, M. Tegmark, M. Zaldarriaga and I. Waga,
Phys. Rev. D 69, 123524 (2004).

\bibitem{Bento04b} M.C. Bento, O. Bertolami and A.A. Sen, Phys. Rev.
D70, 083519 (2004).

\bibitem{Bernardini08} A. E. Bernardini and O. Bertolami, Phys. Rev. D 77, 083506 (2008); Phys. Lett. B 662, 97 (2008).



\end{thebibliography}
\end{document}